\documentclass[pdflatex,sn-mathphys-num]{sn-jnl}

\usepackage{graphicx}%
\usepackage{multirow}%
\usepackage{amsmath,amssymb,amsfonts}%
\usepackage{amsthm}%
\usepackage{mathrsfs}%
\usepackage[title]{appendix}%
\usepackage{xcolor}%
\usepackage{textcomp}%
\usepackage{manyfoot}%
\usepackage{booktabs}%
\usepackage{algorithm}%
\usepackage{algorithmicx}%
\usepackage{algpseudocode}%
\usepackage{listings}%
\usepackage{ulem}	

\newcommand{\erosita}{eROSITA}	
\newcommand{\sco}{Sco X-1}
\newcommand{\art}{ART-XC}
\newcommand{\srg}{{\it SRG}}

\def\spsf{\textit{s}PSF}
\def\arcmin{$'$}

\begin{document}

\title[SRG/ART-XC PSF inflight calibration]{Inflight calibration of SRG/ART-XC point spread function at large off-axis angles}


\author*[1]{\fnm{R.} \sur{Krivonos}}\email{krivonos@cosmos.ru}

\author[1]{\fnm{R.} \sur{Burenin}}

\author[1]{\fnm{E.} \sur{Filippova}}

\author[1]{\fnm{I.} \sur{Lapshov}}

\author[1]{\fnm{A.} \sur{Tkachenko}}

\author[1]{\fnm{A.} \sur{Semena}}

\author[1]{\fnm{I.} \sur{Mereminskiy}}

\author[1]{\fnm{V.} \sur{Arefiev}}

\author[1]{\fnm{A.} \sur{Lutovinov}}

\author[2]{\fnm{B.~D.} \sur{Ramsey}}

\author[2]{\fnm{J.~J.} \sur{Kolodziejczak}}

\author[3]{\fnm{D.~A.} \sur{Swartz}}

\author[3]{\fnm{C.-T.} \sur{Chen}}

\author[2]{\fnm{S.~R.} \sur{Ehlert}}

\author[4]{\fnm{A.} \sur{Vikhlinin}}


\affil*[1]{\orgname{Space Research Institute (IKI), Russian Academy of Sciences}, \orgaddress{\street{Profsoyuznaya 84/32}, \city{Moscow}, \postcode{117997}, \country{Russia}}}

\affil[2]{\orgname{NASA Marshall Space Flight Center}, \orgaddress{ \city{Huntsville}, \postcode{AL 35812}, \country{USA}}}

\affil[3]{\orgname{Science and Technology Institute, Universities Space Research Association}, \orgaddress{\city{Huntsville}, \postcode{AL 35812}, \country{USA}}}

\affil[4]{\orgname{Center for Astrophysics | Harvard \& Smithsonian}, \orgaddress{\street{60 Garden Street} \city{Cambridge}, \postcode{MA 02138}, \country{USA}}}


\abstract{The knowledge of the point spread function (PSF) of the  Mikhail Pavlinsky Astronomical Roentgen Telescope--X-ray Concentrator (\art) telescope aboard the Spectrum-Roentgen-Gamma (\srg) observatory plays an especially crucial role in the detection of point X-ray sources in the all-sky survey and the studies of extended X-ray objects with low surface brightness. In this work, we calibrate the far off-axis shape of the {\art} PSF using in-flight data of {\sco} and the Crab Nebula, in all-sky survey or scan mode, respectively. We demonstrate that the so-called ``slewing'' {\art} PSF (in contrast to the on-axis PSF), in convolution with the detector pixels, is consistent with ground calibration performed at the Marshall Space Flight Center, and can be used to model the PSF up to large off-axis distances in all-sky survey or scan modes. The radial profile of the Crab Nebula in the 4$-$12~keV band shows an extended structure out to $\sim$150'' and is consistent with {\sco} at larger off-axis angles. Finally, we performed an analytic parametrization of the slewing {\art} PSF as a function of energy.
}




\maketitle

\section{Introduction}

The Mikhail Pavlinsky Astronomical Roentgen Telescope--X-ray Concentrator (\art) aboard the Spectrum-Roentgen-Gamma (\srg) orbital observatory\footnote{\url{https://www.srg.cosmos.ru/}} \citep{Sunyaev21}  was launched on July 13, 2019 from the Baikonur Cosmodrome to a halo orbit near the L2 point of the Sun--Earth system. The \srg\ was designed to survey the entire sky in the X-ray energy band. The observatory is equipped with two grazing incidence telescopes: the \erosita\ \citep{Predehl21} and the \art\ \citep{Pavlinsky21}, which operate in overlapping energy bands of 0.2$-$8~keV and 4$-$30~keV, respectively. 

Since the start of operations in July 2019, the \art\ has demonstrated stable in-flight performance consistent with pre-flight expectations. A number of X-ray surveys have been conducted by the \art\ during the \srg\ Calibration and Performance Verification (Cal-PV) phase: the Galactic Plane Survey near Galactic Longitude $L\simeq20^{\circ}$ \citep{2023AstL...49..662K} and the Galactic Bulge deep survey \citep{2024MNRAS.529..941S}. After completion of the first two all-sky scans between December 2019 and December 2020, \cite{2022A&A...661A..38P} compiled a catalog of sources detected by \art\ in the 4$-$12\,keV energy band. An updated version of this catalog, based on the entire data set accumulated by \art\ during the first ${\sim}4.4$ all-sky surveys is presented by \cite{2024A&A...687A.183S}. A number of studies of individual point-like and extended X-ray sources have also been performed, e.g., of the young compact cluster of massive stars Westerlund 2 \citep{2023MNRAS.525.1553B}; of several X-ray pulsars \citep{2021ApJ...912...17L,2022A&A...661A..45T,2022A&A...661A..21D,2022A&A...661A..28L,2022A&A...661A..33M,2022AstL...48..798G,2023AstL...49..240S}; and of the supernova remnant Puppis A \citep{2022MNRAS.510.3113K}. The detailed properties of the PSF at large radii are important for modeling the extended emission of the X-ray-bright Coma and Ophiuhus galaxy clusters; and for removing contribution of bright X-ray sources from the map of the inner Galactic Center. 

Knowledge of the angular response of the \art\ optics at large off-axis angles is important not only to study the extended sources, but also to have better sensitivity for detection of weak point sources, especially during the scanning observations \citep{2024A&A...687A.183S,2024MNRAS.529..941S}. The point spread function (PSF) of the \art\ is characterized by the relatively wide wings, formed by  single-reflected photons falling onto the detectors with angular offsets up to ${\sim}50'$ from the optical axis \citep{Pavlinsky21}. The analysis of ground calibrations of the \art\ mirror modules have concentrated on the central part of the PSF, providing only analysis out to $24'$ for one selected mirror module \citep{2017ExA....44..147K}, i.e. lacking the calibration of PSF at far off-axis angles. In this work we present the results of the \art\ PSF inflight calibration at large off-axis angles using the wide-field observations of the bright X-ray sources \sco\ and the Crab Nebula.

The paper is organized as follows. In Sect.~\ref{sec:psf} we briefly summarize properties of the \art\ optics. We reanalyze data obtained at the Marshall Space Flight Center (MSFC) to slightly extend the angular coverage of the ground PSF measurements, as described in Sect.~\ref{sec:msfc}. The PSF in-flight calibration (Sect.~\ref{sec:inflight}) is based on observations of \sco\ and the Crab Nebula. We discuss and summarize the results, respectively, in Sect.~\ref{sec:discussion} and Sect.~\ref{sec:summary}. 

\section{\art\ mirror modules}
\label{sec:psf}

The \art\  consists of seven independent, co-aligned, X-ray telescopes, each containing a Wolter Type-I mirror system (MS) and a position-sensitive focal plane detector. \cite{Pavlinsky21} presents a detailed description of the telescope's systems. Here we only outline the basic properties of the X-ray optics. Each mirror module consists of 28 concentrically nested grazing-incidence paraboloid and hyperboloid mirror pairs. Each of the \art\ mirrors is coated with iridium (Ir) to enhance its  reflectivity up to 30 keV. The nominal focal length of the \art\ mirror modules is 2700~mm, however, the detectors are installed with an intra-focal 7~mm defocusing to provide more uniform angular resolution across the field of view (FOV). The FOV for double reflected photons (reflected first by parabolic and then by hyperbolic mirror) is 36 arcmin in diameter, which is mainly determined by the working area of the CdTe semiconductor double-sided strip detectors with pitch equivalent to $45''$ \citep{2014SPIE.9144E..13L}. The \art\ on-axis PSF is characterized by a `core' with ${\sim}30-35''$  half-power diameter (HPD, corresponding to half of the focused X-rays), that is comparable to the angular size of the detector pixel. A wider component of the \art\ PSF, typical for Wolter Type-I optics, is referred to as a `wing' visible at large off-axis angles. The PSF core and wing components can be represented in analytical form as a combination of two King's functions \citep[see below and][]{2017ExA....44..147K}.

Photons that only undergo a single reflection inside the optics, can reach detector from angular scales up to ${\sim}50'$ off-axis. These photons do not form a direct image \citep{2019ExA....48..233P} but, in all-sky survey or scan modes, can form a parasitic `halo' on the sky image from bright sources that transit at a range of azimuthal locations and off-axis angles.

We stress, however, that the bulk of the flux reaching the detector is from doubly-reflected (focused) X-radiation. The characterization of the integrated shape of singly-reflected photons at large off-axis angles is mainly needed to take this parasitic halo into account as a component of the extended X-ray sky surface brightness.

\section{Ground calibration of PSF at MSFC}
\label{sec:msfc}
The ground calibration of eight \art\ mirror modules (ARTM1-8) has been performed using the 104-m X-ray beam facility at the Marshall Space Flight Center (MSFC).  The X-radiation source was a copper anode of the X-ray tube with applied 12 kilovolt voltage. The spectrum of the X-ray source is dominated by copper (Cu) K$\alpha$ (8.04 keV) and K$\beta$ (8.9 keV) lines, thus, generating X-ray beam dominated by line emission at energy ${\sim}8$~keV with a contribution from the continuum. The shape of the PSF was measured with a high-resolution CCD camera installed in the focal plane with defocusing of 7~mm, as required by the {\art} design. \cite{2017ExA....44..147K} performed a characterization of the PSF for each \art\ mirror module at various angular distances, considering only the central part of the PSF within $700''$ from the optical axis. In this work, we reanalyzed the MSFC calibration data, with the aim to extend the PSF measurements to the full $\sim2000''$ size of the CCD camera. 

\begin{figure}
	\centerline{\includegraphics[width=\columnwidth]{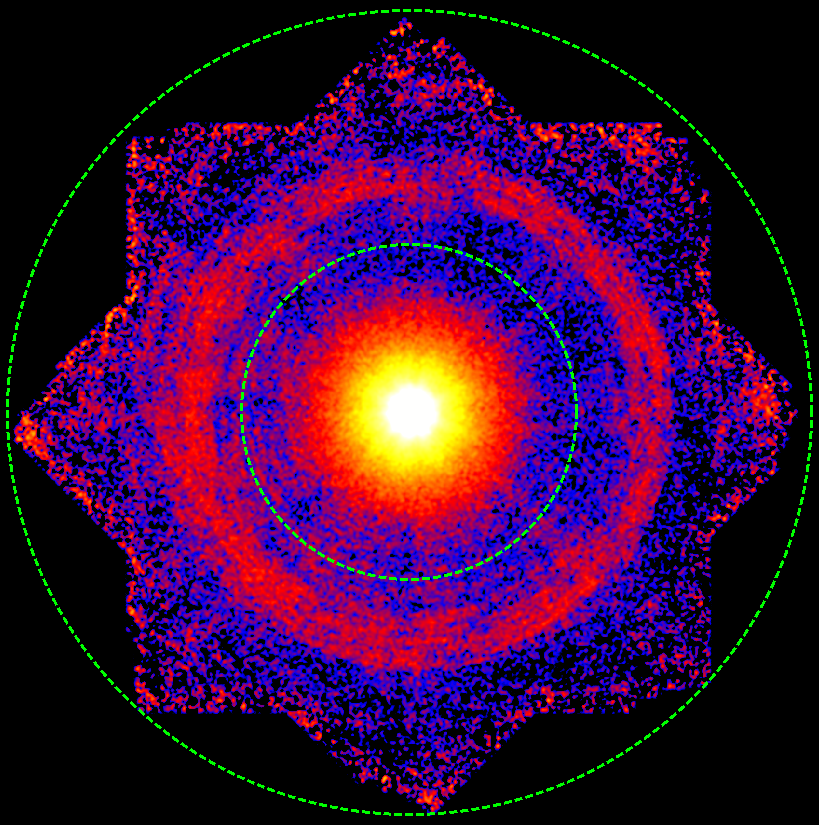}}
    \caption{On-axis \art\ PSF from ground calibration, averaged over all flight modules. The radius of the outer green dashed circle is $R=28'$ from the optical axis. The region internal to the inner circle with $R=700''$ ($R\simeq11.7'$) was originally considered by \citealt{2017ExA....44..147K} to calibrate the central part of the PSF. The broad red ring outside the inner circle is caused by X-radiation singly-reflected from the innermost hyperboloid shell due to a gap between that shell and an inner opaque baffle used only during calibration. The gap is not present in the flight configuration.}
    \label{fig:onaxis}
\end{figure}

\begin{figure}
\centerline{\includegraphics[width=\columnwidth]{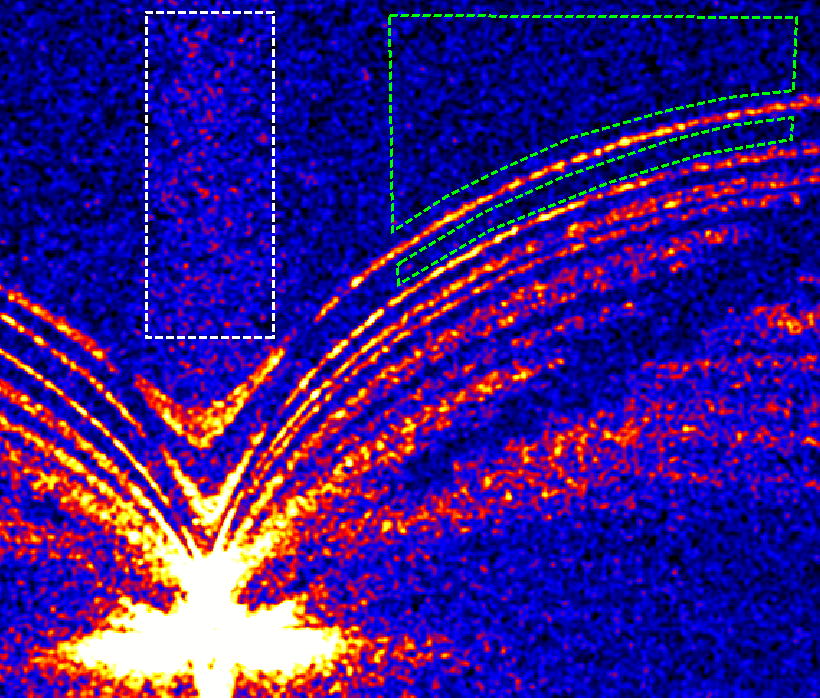}}
\caption{An example of the far off-axis ($\theta=18'$) \art\ PSF image measured with ARTM1. The green dashed polygons show regions used for the estimation of the internal CCD background. The white dashed rectangle denotes the approximate region of the CCD image contaminated by readout effects caused by the very bright X-ray beam.}
\label{fig:example}
\end{figure}

In Fig.~\ref{fig:onaxis} we show the on-axis PSF image, averaged over all \art\ flight mirror modules ARTM1-5 and ARTM7-8, measured in the ground calibration. Hereafter, we refer to this averaged PSF as ARTM0 (ARTM6 has been selected as a spare flight unit and has been reserved for ground tests). One can notice the bright central core and a ring-like structure, produced by photons singly-reflected from the innermost hyperboloid mirror shell caused by a gap between the inner shell and a central baffle used to prevent light from passing directly through the center of the mirror assembly.  In contrast to  \cite{2017ExA....44..147K}, who used standard CCD dark-current subtraction procedure, here we estimate the background from the actual CCD images. For near on-axis PSF measurements at off-axis angles $\Theta<9'$, we estimated the internal background near the edges of the CCD matrix where we assume the PSF wings are not significant. In the case of far off-axis calibrations, where the PSF fills nearly the whole CCD frame, we defined background regions outside of or between the streaks of individual mirror's shells as demonstrated in Fig.~\ref{fig:example}. Such a selection of background regions is motivated by the fact that X-ray reflectivity of the mirrors is dominated by geometry (see Sec.~\ref{sec:discussion:energy}), and additional scattering of photons caused by mirrror's roughness is not significant. Finally, we have also removed bright CCD regions caused by readout effects, as shown by the rectangle in Fig.~\ref{fig:example}.

\begin{figure}
\centerline{\includegraphics[width=\columnwidth]{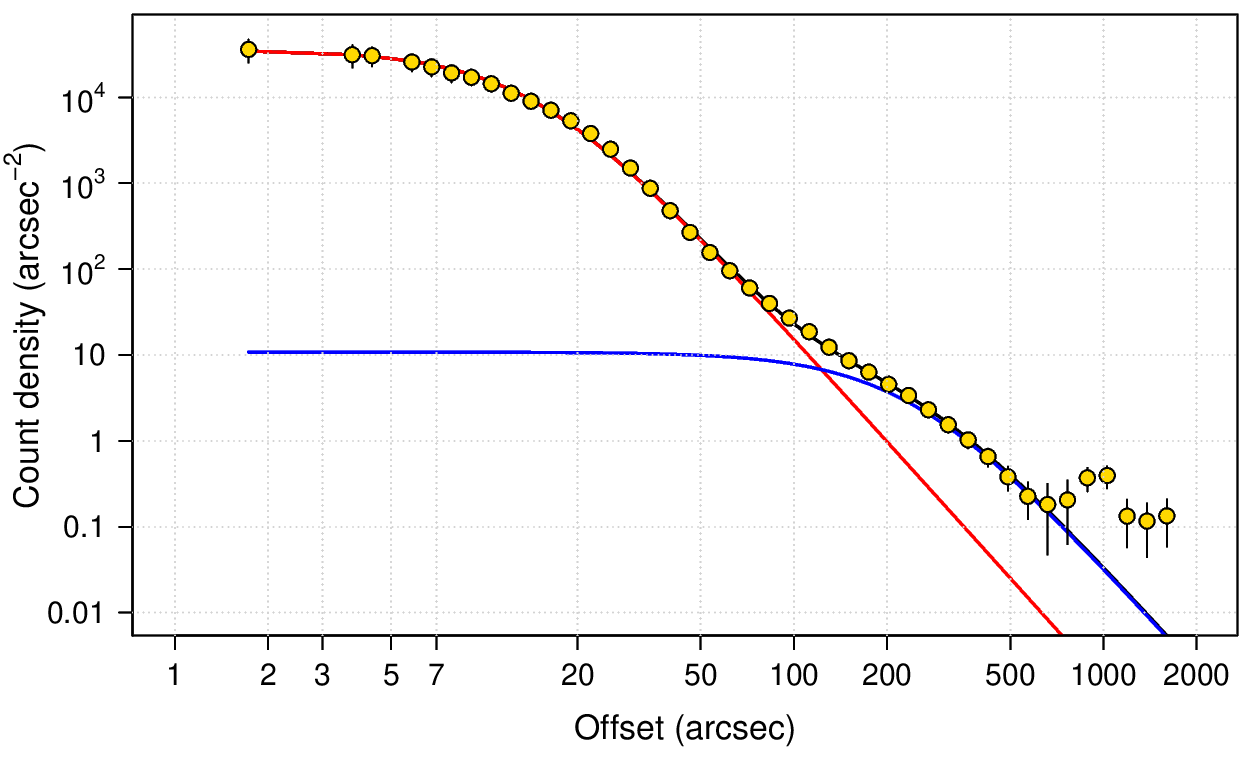}}
\caption{Radial profile of the on-axis PSF measured in ground calibration and averaged over all \art\ mirror flight modules. The King proﬁles for the PSF core and wing are shown with solid red and blue lines, respectively. Note the enhancement at $\sim1000''$ due to singly-reflected photons.}
    \label{fig:king}
\end{figure}

Fig.~\ref{fig:king} shows the radial profile of the on-axis PSF, averaged for all flight \art\ mirror modules (ARTM0). Following \cite{2017ExA....44..147K} we model its narrow PSF core and wide wing with King's function \citep{1962AJ.....67..471K}:
\begin{equation}
K(r,\sigma,\gamma) = \left(1+\frac{r^{2}}{\sigma^{2}}\right)^{-\gamma}
\label{eq:king}
\end{equation}
The parameter $\sigma$ is the characteristic size of the core, $\gamma=2$ determines the weight of the tails, and $r$ is the off-axis angle in arcsec. The linear combination of two King functions represents the radial profile of the \art\ on-axis PSF:
\begin{equation}
\begin{split}
PSF(r) = &~N_{\rm core} \times \big[  f_{\rm core} K(r,\sigma_{\rm
  core},\gamma_{\rm core})\\ 
&+ f_{\rm wing} K(r,\sigma_{\rm wing},\gamma_{\rm wing})\big],
\end{split}
\label{eq:psfform}
\end{equation}
where $N_{\rm core}$ is an arbitrary overall normalization factor and $f_{\rm core}+f_{\rm wing}=1$. The best-fitting values of the free parameters are $N_{\rm core}=(1.42\pm0.13)\times10^{-3}$~arcsec$^{-2}$, $\sigma_{\rm core}=14.4\pm0.4$~arcsec, $f_{\rm wing}=(3.0\pm0.6)\times10^{-4}$, and $\sigma_{\rm wing}=239\pm22$~arcsec. Note that we have renormalized $N_{\rm core}$ so that
\begin{equation}
\int_{\rm FOV} PSF(r) d\Omega = \int_0^{2\pi} \int_0^{\infty} PSF(r)rdrd\phi = 1.
\end{equation}

\begin{figure}
\centerline{\includegraphics[width=\columnwidth]{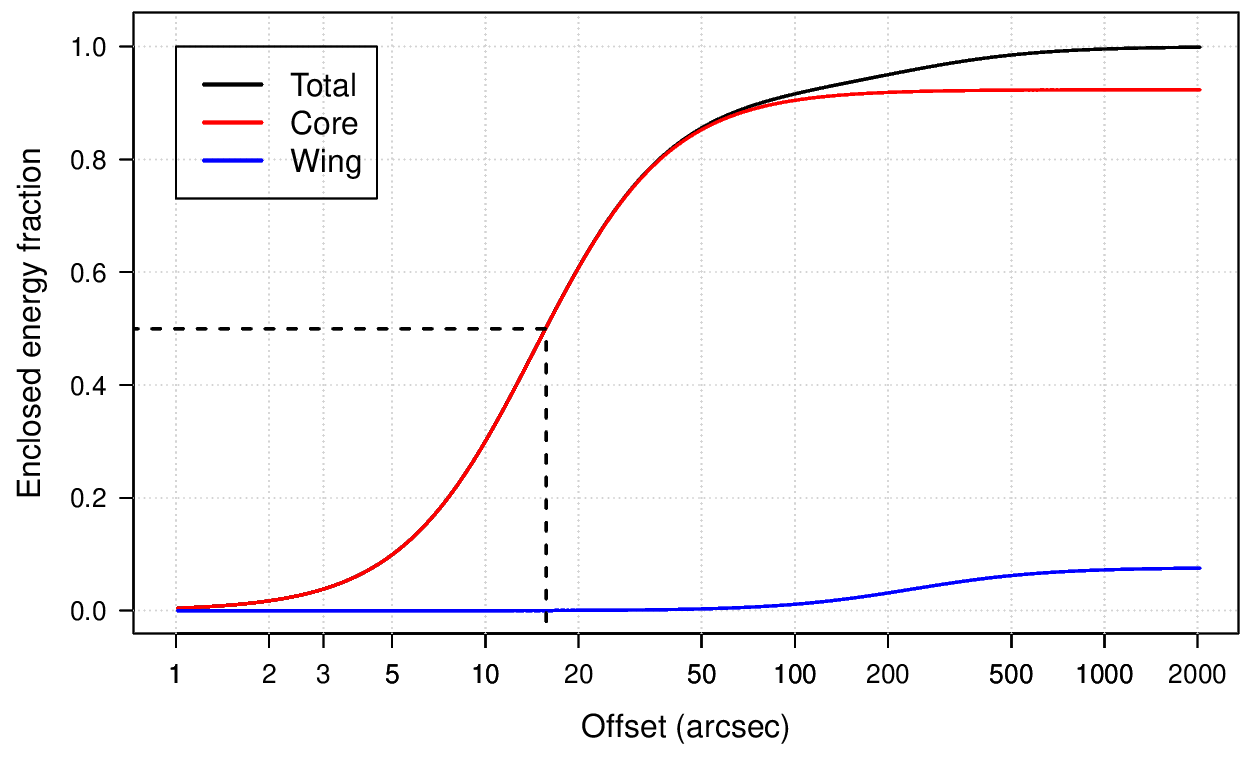}}
\caption{Enclosed energy fraction of the on-axis PSF (ARTM0), derived from the analytic form Eq.~\ref{eq:king}-\ref{eq:psfform} for the PSF core (red) and wing (blue) components, and total (black). Half of the total EEF, corresponding to the HPD value, is shown by the dashed line (HPD$_{\rm total}=31.5$~arcsec).}
\label{fig:eef}
\end{figure}

Using the analytic form (Eqs.~\ref{eq:king}, \ref{eq:psfform}), we calculated the enclosed energy fraction (EEF) as a function of off-axis angle $r$, shown in Fig.~\ref{fig:eef}. The HPD for the King functions (Eq.~\ref{eq:king}) with $\gamma=2$ is reached at an off-axis angle $r=\sigma$, so that the HPD corresponds to $2\times\sigma$. For the two-component PSF, the presence of wings increases the total HPD by ${\sim}10\%$, i.e., HPD$_{\rm core}=2\times\sigma_{\rm core}=28.8\pm0.4$~arcsec and HPD$_{\rm total}=31.5$~arcsec. The \art\ on-axis PSF can also be characterized by the parameter $W_{90}=156$~arcsec (the 90\% enclosed energy diameter), and by the ${\sim}7\%$ contribution of the wing component to the total EEF calculated at $r=\infty$.

\section{Inflight calibration}
\label{sec:inflight}

MSFC ground calibration of the \art\ mirror modules with high-resolution CCD matrix described in \cite{2017ExA....44..147K} and in Sect.~\ref{sec:msfc} is essential for characterization of the angular response of the X-ray optics. However, the knowledge of the PSF in combination with the physical (flight) detector is of practical interest, mainly due to the fact that the angular size of the flight detector pixel ($45''$) is comparable to the HPD of the optics (HPD$_{\rm total}=31.5''$, see above). Saying that, one can draw a conclusion that the most stable optical response of the \art\ is reached in slewing; when the narrow core of the PSF is averaged within the detector pixel. This is achieved during observations in scan and in all-sky survey modes \citep{Sunyaev21}. Additionally, the {\art} image reconstruction procedure includes a pixel-randomization step which randomizes the photon coordinates  within the physical pixel to avoid creating artificial structures in X-ray images.

\begin{figure}
\centerline{\includegraphics[width=\columnwidth]{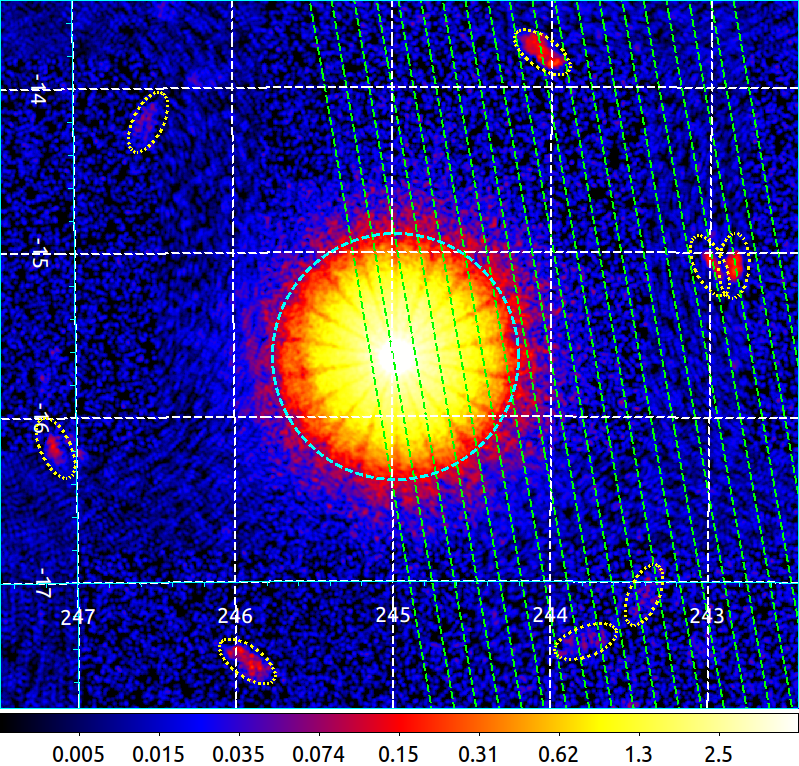}}
\caption{\srg/\art\ sky image of Sco X-1 (in the center) obtained in the 4$-$12~keV band in all-sky survey mode. The exposure-corrected image is shown in logarithmic scale in the range from zero to 0.15\% of the peak value. The units of the image are counts~s$^{-1}$~pix$^{-1}$. The image is smoothed with a Gaussian kernel ($\sigma=30''$). The dashed cyan circle denotes the angular offset of $2700''$ from the Sco X-1 position. Ellipses are known artifacts, caused by light leaks of the \art\ structure. Radial spoke-like features are due to ribs of MS supports. There are 9 supports in each MS and each MS is rotated 60$^{\circ}$ from its neighbor giving the appearance of 18 spokes. Green dashed lines demonstrate typical FOV passages during the all-sky survey, characterized by ${\sim}8'$ step between passages (at the position of {\sco}) and slewing speed of $1.5'$~s$^{-1}$. The currently shown passages correspond to four-day block of sky slewing observations (ObsID 11000600100, see Table~\ref{tab:obs}).}
\label{fig:sco}
\end{figure}

In-flight wide-field scanning observations of a bright source allows us to characterise the central part of the \art\ PSF and its wide wing in slewing mode, hereafter referred to as a slewing PSF (\spsf). The \art\ (\spsf) is in fact a result of a convolution of the PSF with a physical detector pixel. Note that the calibration of the {\spsf} in a laboratory is a difficult task due to a wide range of angular offsets and azimuthal directions required to set up in a continuous (non-discrete) way. Below we present measurements of the {\spsf} in 4$-$12~keV energy band using bright X-ray sources {\sco} in \srg\ all-sky survey mode and the Crab Nebula in scanning mode. We  also note that the shape of the {\spsf} obtained in this way depends on the scanning mode parameters determined mainly by the slew speed and the step between the passages of the source within the FOV, because the resulting {\spsf} is a result of integration of the \art\ PSF, modified by the vignetting function at a given moment of time. 

\begin{figure}
\centerline{\includegraphics[width=\columnwidth]{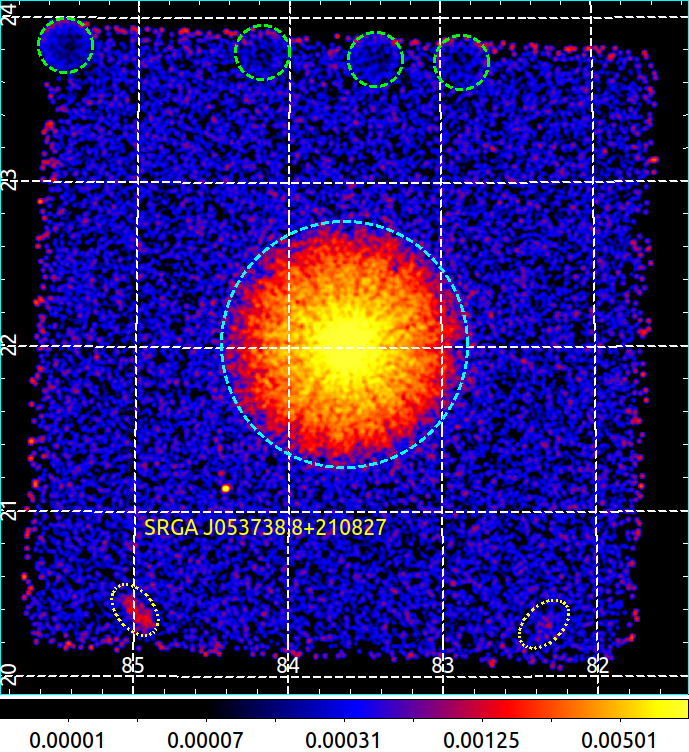}}
\caption{\srg/\art\ $3.5^{\circ}{\times}3.5^{\circ}$ scanning observation of the Crab Nebula (in the center), obtained in the 4$-$12~keV band. The units of the image are counts~s$^{-1}$~pix$^{-1}$. The image is shown in logarithmic scale in the range from zero to 0.1\% of the peak value. The image is smoothed with a Gaussian kernel ($\sigma=30''$). The dashed cyan circle denotes an angular offset of $2700''$ from the Crab position. Green dashed circles ($20'$ in diameter) show {\srg} technological pointing observations, necessary to stabilize the spacecraft (``scan parkings''). Ellipses are known artifacts, caused by light leaks of the \art\ structure. The position of X-ray source SRGA J053738.8+210827 (zet Tau, a gamma Cas star) from ARTSS \citep{2024A&A...687A.183S} is labeled.}
\label{fig:crab}
\end{figure}

\begin{figure}
\centerline{\includegraphics[width=\columnwidth]{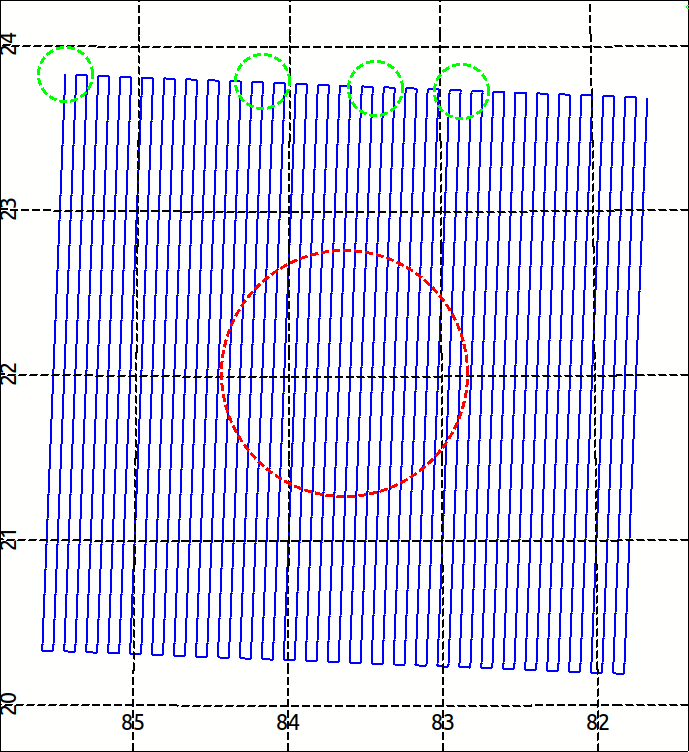}}
\caption{Blue lines visualize the scheme of FOV passages during the \srg/\art\ $3.5^{\circ}{\times}3.5^{\circ}$ scanning observation of the Crab Nebula (see Fig.~\ref{fig:crab} for reference). This scanning observation is characterized by $4'$ step between passages, slewing speed of $0.15'$~s$^{-1}$ and total exposure of 22 hours. Similar to Fig.~\ref{fig:crab}, the dashed red circle denotes an angular offset of $2700''$ from the Crab position and green circles show scan parking positions.}
\label{fig:crab:scan}
\end{figure}

{\sco} has been observed between December 12, 2019, and March 7, 2022, during the on-going {\art} all-sky survey \citep[ARTSS,][]{2022A&A...661A..38P,2024A&A...687A.183S} covered by eight blocks of typical four-day sky slewing observations listed in Table~\ref{tab:obs}. Fig.~\ref{fig:sco} demonstrates the exposure-corrected sky image of {\sco} in the 4$-$12~keV band. To avoid spatial confusion, we only show FOV passages for the first slewing block with ObsID 11000600100. The total effective exposure in any point of the image is about 190~s. The Crab Nebula was observed with \art\ in a $3.5^{\circ}{\times}3.5^{\circ}$ scanning  mode observation on 18 March 2024 (Table~\ref{tab:obs}) with a total exposure of 22 hours. Fig.~\ref{fig:crab} shows the Crab exposure-corrected sky image in the 4$-12$~keV energy band, summed over all seven {\art} modules.  The effective exposure of the image is about 110~s. The FOV passages during the scan are shown in Fig.~\ref{fig:crab:scan}. Finally, we note that for PSF calibration we utilize a FOV with a diameter of $20'$ (in contrast to a full FOV $36'$ in diameter).

\begin{table}
    \centering
    \begin{tabular}{rrc}
\hline
 Date/Time Begin & Date/Time End & ObsID \\
\hline
\multicolumn{3}{c}{\sco: all-sky survey}\\
02 Mar 2020 22:00 &	06 Mar 2020 22:00 &	11000600100 \\ 	
06 Mar 2020 22:00 &	10 Mar 2020 22:00 &	11000600200 \\ 	
02 Sep 2020 19:00 &	07 Sep 2020 19:00 &	12000300100 \\ 	
18 Feb 2021 15:30 &	22 Feb 2021 15:30 &	13000200500 \\ 	
22 Feb 2021 15:30 &	25 Feb 2021 15:30 &	13000200600 \\ 	
27 Aug 2021 21:00 &	01 Sep 2021 21:00 &	14000200600 \\ 	
22 Feb 2022 20:30 &	27 Feb 2022 20:30 &	15000200500 \\ 	
27 Feb 2022 20:30 &	02 Mar 2022 20:30 &	15000200600 \\
\hline
\multicolumn{3}{c}{Crab Nebula: scan mode}\\
18 Mar 2024 02:37 & 19 Mar 2024 00:24 & 02420000319 \\
\hline
\bottomrule
\end{tabular}
    \caption{The list of {\srg}/{\art} observations used for in-flight PSF calibration.}
    \label{tab:obs}
\end{table}

\begin{figure}
    \centering
    \includegraphics[width=\columnwidth]{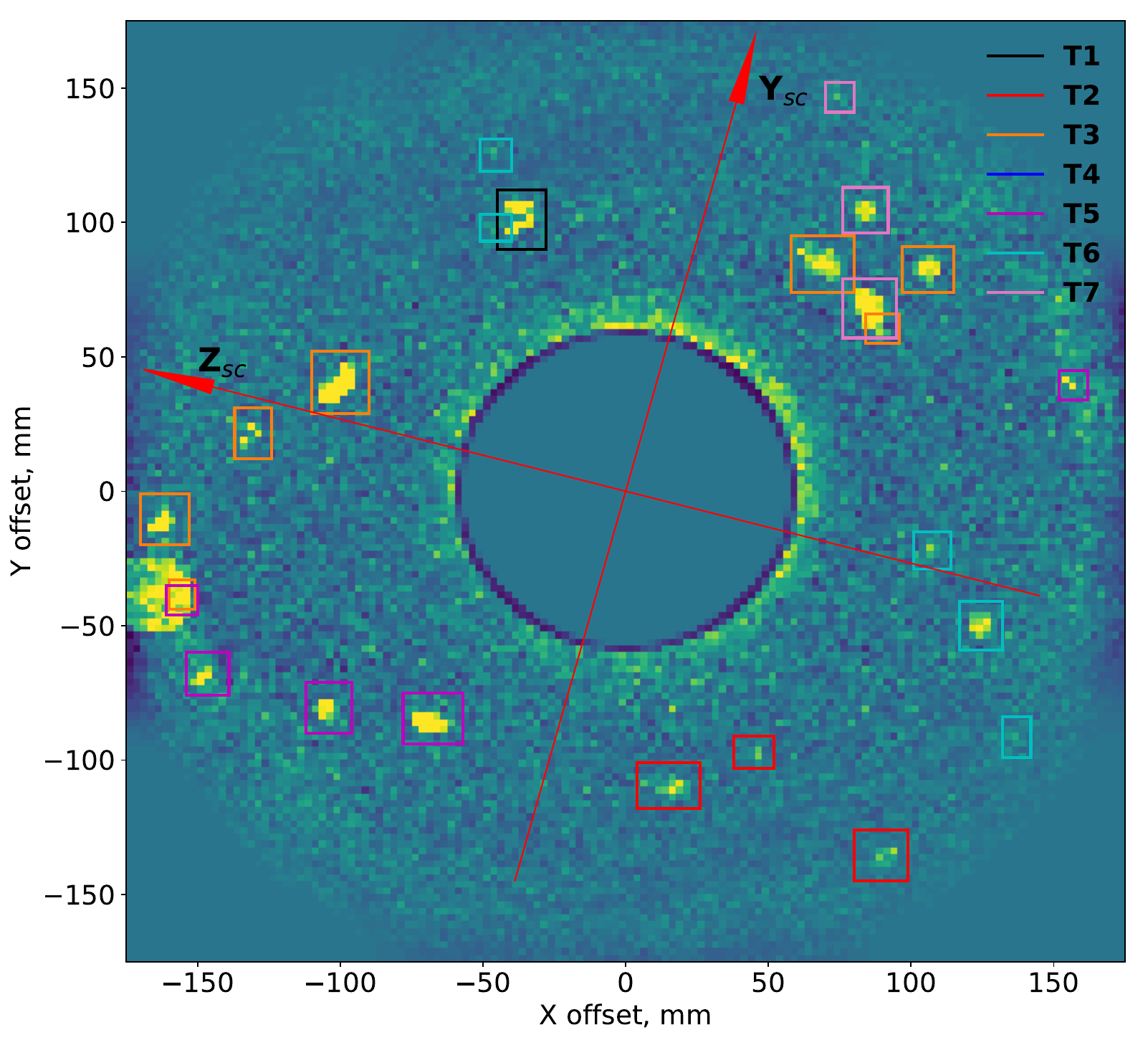}
    \caption{Surface brightness of the {\sco} deprojected on the mirror system mounting plate around each detector. The rectangles of different colors mark observed stray light for different detectors. As one can see, the stray light patterns are unique to each telescope module, and telescope T4 has no stray light leaks at all. }
    \label{fig:stray_light}
\end{figure}

The artifacts, seen at 1$^\circ$ to 1.6$^\circ$ offsets from bright sources (Figs.~\ref{fig:sco} and \ref{fig:crab}) are the product of stray light leaks, created by partial transparency of the mirror system mounting plate in a few spots. Each detector is shielded from  stray light with a 30~cm high collimator that blocks light from angles larger than 1.6$^\circ$ \citep{Pavlinsky21} and the mirror system itself blocks light from within 24\arcmin\ from the center of the detector. Other angles are covered by an opaque mounting plate which however have few semi-opaque spots within 16~cm radius around each mirror system. Using scanning observations of  Sco~X-1, we deprojected all events in the plane of the mounting plate, located at 3~m distance above the detectors, which revealed patterns of light leaks at various offsets, corresponding to the partially transparent spots in the mounting plate (Fig.~\ref{fig:stray_light}). The location of such spots were described by rectangular regions in the deprojected coordinates, which are used to filter out stray light in scientific observations and are listed in  Table~\ref{tab:liks_patches}. As one can see, patterns of the stray light leaks are unique to each telescope and the central telescope (T4) does not have stray light leaks at all.

\begin{table}
    \centering
    \begin{tabular}{lcrrrr}
\hline
 N & T & X$_{\rm low}$ & X$_{\rm high}$ & Y$_{\rm low}$ & Y$_{\rm high}$ \\
 & & mm & mm & mm & mm \\
\hline
1 & 1 & -42 & -32 & 93 & 108 \\
2 & 2 & 7 & 22 & -115 & -105 \\
3 & 2 & 41 & 48 & -100 & -95 \\
4 & 2 & 83 & 95 & -142 & -130 \\
5 & 3 & -167 & -157 & -17 & -5 \\
6 & 3 & -134 & -128 & 15 & 27 \\
7 & 3 & -107 & -94 & 32 & 48 \\
8 & 3 & 61 & 76 & 77 & 91 \\
9 & 3 & 100 & 111 & 77 & 87 \\
10 & 3 & -157 & -155 & -41 & -37 \\
11 & 3 & 87 & 92 & 58 & 62 \\
12 & 5 & -158 & -154 & -43 & -39 \\
13 & 5 & -151 & -143 & -73 & -64 \\
14 & 5 & -109 & -100 & -87 & -75 \\
15 & 5 & -75 & -61 & -91 & -79 \\
16 & 5 & 155 & 158 & 37 & 41 \\
17 & 6 & 104 & 110 & -26 & -19 \\
18 & 6 & 120 & 128 & -56 & -45 \\
19 & 6 & 135 & 138 & -96 & -88 \\
20 & 6 & -48 & -44 & 122 & 127 \\
21 & 6 & -48 & -44 & 96 & 99 \\
22 & 7 & 73 & 76 & 144 & 148 \\
23 & 7 & 79 & 88 & 99 & 109 \\
24 & 7 & 79 & 91 & 60 & 75 \\
\hline
\bottomrule
\end{tabular}
    \caption{The list of rectangular patches covering the stray light leaks from the bright sources in the detector coordinate system, deprojected on to the mirror system mounting plane. Second column denotes inflight numbering of the {\art} telescope modules.}
    \label{tab:liks_patches}
\end{table}

\begin{figure}
\centerline{\includegraphics[width=\columnwidth]{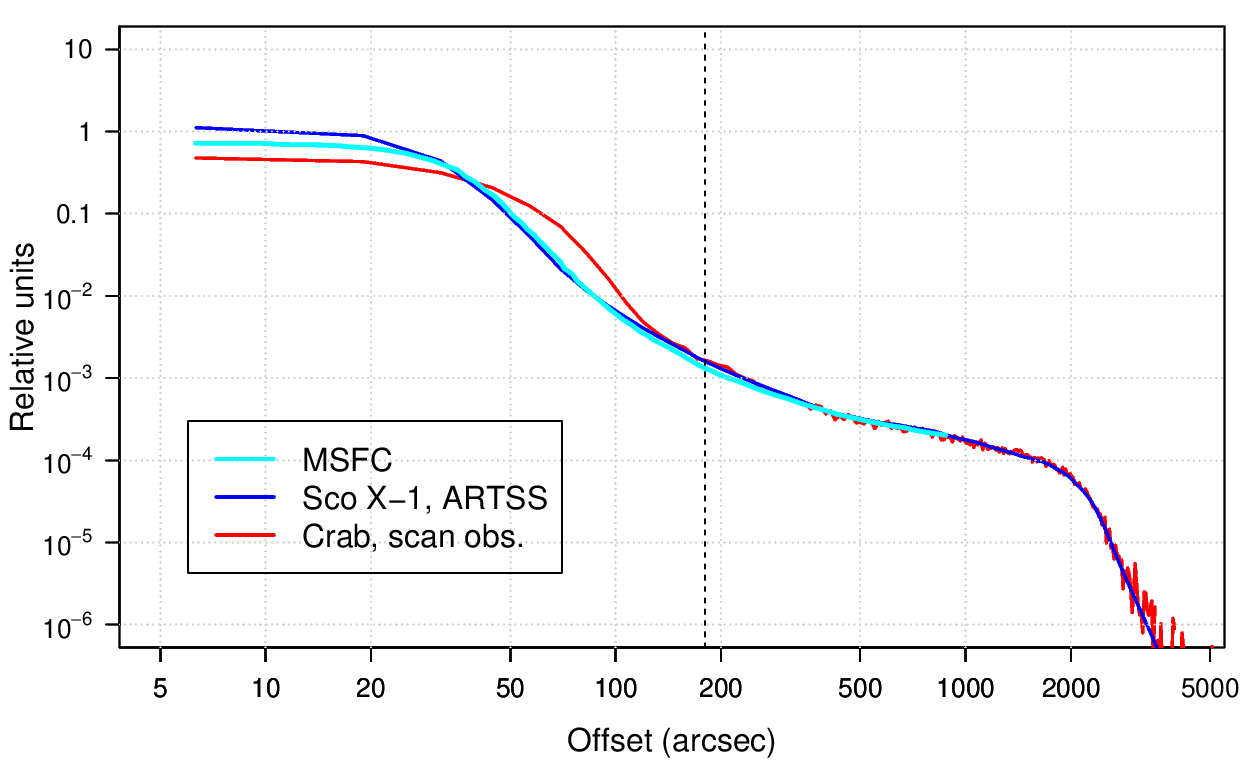}}
\caption{Radial surface brightness profiles of the {\art} {\spsf} (4$-$12~keV), based on {\sco} and Crab Nebula observations in all-sky survey and scan modes, respectively. Cyan line shows MSFC PSF averaged over different offsets and azimuthal angles, and weighted by the \art\ vignetting function. All radial profiles are normalized within $3'$ offset, as shown by vertical dashed line.}
\label{fig:spsf}
\end{figure}

In Fig.~\ref{fig:spsf} we show the resulting {\art} {\spsf} based on ground and inflight data. The MSFC {\spsf} has been obtained by averaging ${\sim}8$~keV ARTM0 PSF (Sect.~\ref{sec:msfc}) over different offsets and azimuthal angles, and weighted by the  {\art} vignetting function. Inflight {\spsf} radial profiles of {\sco} and Crab has been directly extracted from 4$-$12~keV background-subtracted sky images. For relative comparison, all radial profiles have been normalized within $3'$ offset. As seen from the figure, {\sco} is consistent with a point source, since it follows the MSFC {\spsf}, however, the very central part deviates slightly compared to the ground measurements. To explain this discrepancy, one might assume that it is a combination of finite source distance\footnote{The finite distance reduces the area of the optic illuminated with X-rays that reach the focal plane in two reflections thus essentially suppressing the effective area.} and gravity sag in the case of the ground MSFC measurements and perhaps a contribution from the large pixel size in the in-flight measurement. Note that the \art\ detectors show high efficiency even for very bright sources \citep[see e.g.,][]{2024MNRAS.531.4893M}. The count rates were corrected for {\art} dead-time of 770 $\mu s$ \citep[][]{Pavlinsky21}. The 4$-$12~keV Crab radial profile (Fig.~\ref{fig:spsf}) shows the extended nature of the source, although {\art} is not capable of resolving its details.

\section{Discussion}
\label{sec:discussion}

In the working energy band 4$-$12~keV of the {\art}, where the X-ray mirror's effective area is maximum \citep{Pavlinsky21}, the shape of the {\art} {\spsf} is of critical interest for determining the properties of the X-ray mirror system at large off-axis angles.

The comparison of {\spsf} radial profiles, obtained in three different ways and in different energy bands is shown in Fig.~\ref{fig:spsf}. As seen from the figure, MSFC ground-based {\spsf} demonstrates a good agreement with inflight {\sco} radial profile\footnote{The wings of the PSF at large distances from the PSF peak are due to small-scale scattering from the optical surface and that does not change between ground measurements and on-orbit. Therefore, the fact that the wings have the same shape in flight and on the ground is as expected.} at offsets larger than ${\sim}20''$; and the extended structure of Crab makes full agreement between all measurements only possible at off-axis angles larger than ${\sim}150''$.

As mentioned above, the large-scale {\art} parasitic PSF halo is formed by single-reflected photons in all-sky survey or scan modes. The shape of the halo, usually appeared around bright X-ray sources, depends on the scanning mode parameters, e.g., the slew speed and the step between the passages of the source within the FOV. Additionally, the strong change in source intensity during far off-axis passages can significantly enhance or suppress registered single-reflected photons, producing an asymmetric halo on the final sky image.

\subsection{{\spsf} enclosed energy fraction}
\label{sec:discussion:eef}

\begin{figure}
\centerline{\includegraphics[width=\columnwidth]{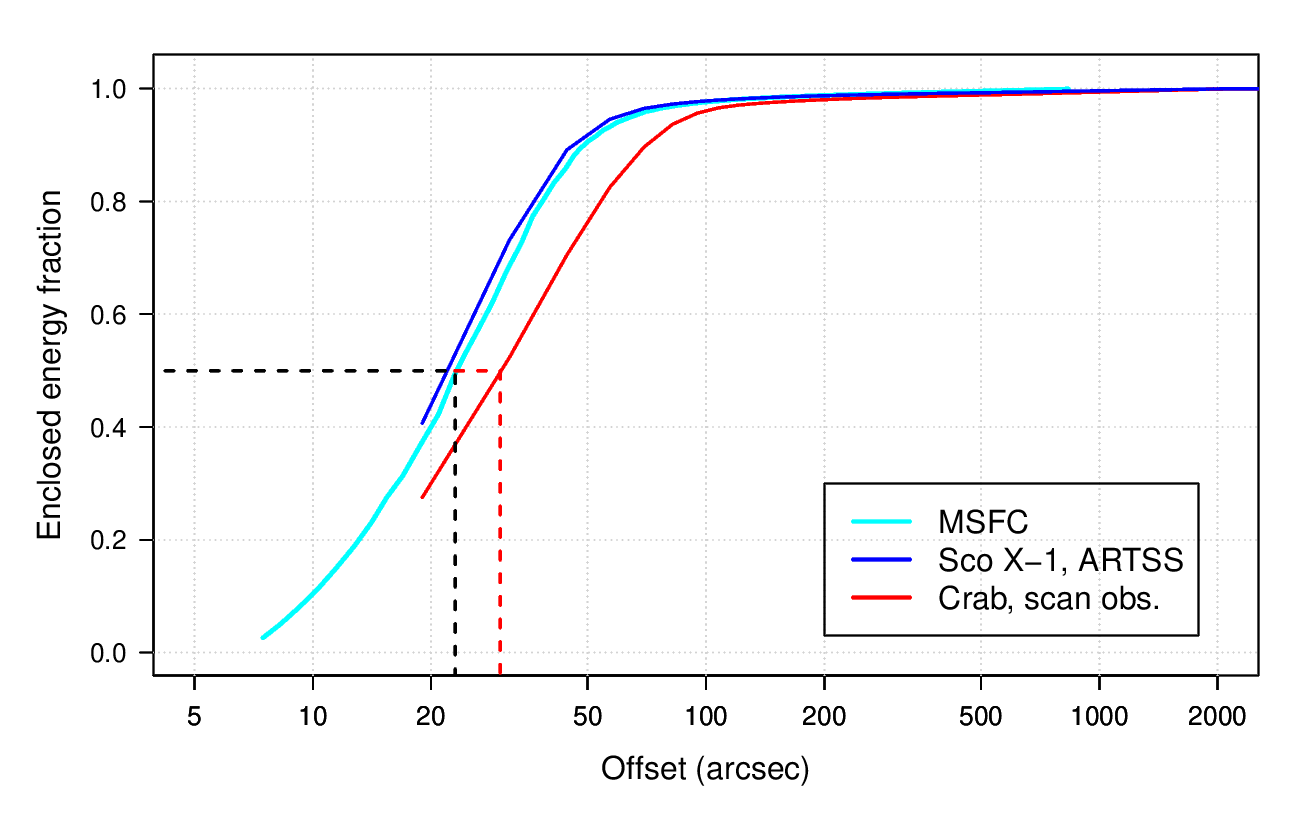}}
\caption{Enclosed energy fraction of the \art\ PSF in scan mode (\spsf, 4$-$12~keV), derived from MSFC ground calibrations (ARTM0) the inflight observations of Sco X-1 and Crab Nebula. Dashed black and red lines show HPD values of $48''$ and $60''$, respectively.}
\label{fig:eef_spfs}
\end{figure}

The enclosed energy fraction (EEF) measures the fraction of the light enclosed within a given radius. This important PSF metric characterizes the collective power of a telescope's optics as a function of angular distance from the source position. In Figure~\ref{fig:eef_spfs} we show the EEF of the \art\ {\spsf} for the three cases studied here. Despite somewhat of a discrepancy between ground measurements and {\sco} at small off-axis angles, their EEFs are in a good agreement. The estimated HPD of the {\spsf} is about $48''$. As expected, the Crab Nebula EEF is shifted to a higher HPD value of $60''$ because it is an extended source.

Taking into account all of the above, we conclude that the measurement of the slewing PSF with {\sco} is consistent with ground calibration at MSFC and can be used to model the {\art} {\spsf}  up to large off-axis distances. 

\subsection{{\spsf} energy dependence}
\label{sec:discussion:energy}

The X-ray reflectivity of the \art\  mirrors decreases with energy above the Ir L-edges, which is manifest by a drop  in the effective area above about 12~keV, followed by a slow decline at higher energies \citep{Pavlinsky21}. As expected, the spatial extent of the {\spsf} will demonstrate shrinkage due to ``cutoff'' of the outer mirror shells that do not contribute substantial area at high energies.

\begin{figure}
\centerline{\includegraphics[width=\columnwidth]{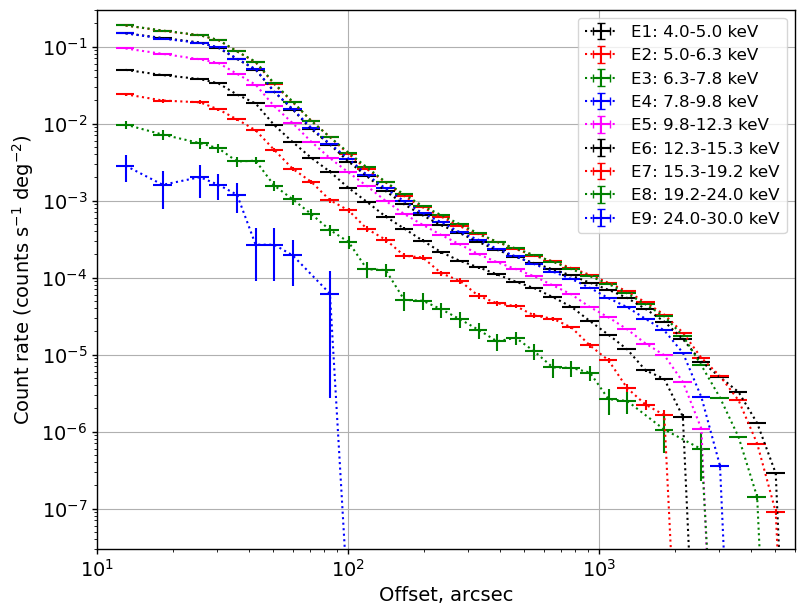}}
\caption{Background-subtracted radial surface-brightness profiles of {\sco} in different energy bands. Note that non-significant bins ($S/N<1\sigma$) are ignored for simplicity.}
\label{fig:scox1}
\end{figure}

\begin{figure}
\centerline{\includegraphics[width=\columnwidth]{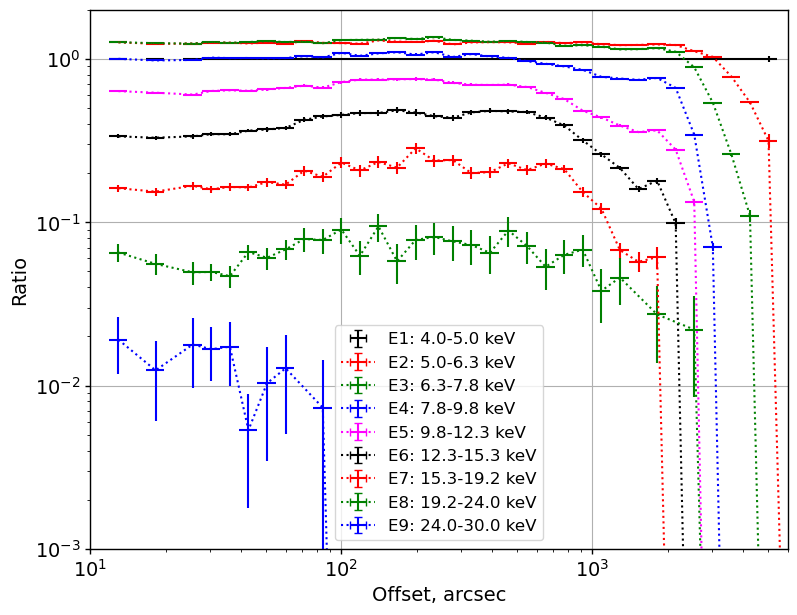}}
\caption{E2$-$E9 radial surface-brightness profiles (Fig.~\ref{fig:scox1}), shown as a ratio with respect to radial profile in 4.0$-$5.0~keV (E1) energy band. }
\label{fig:scox1:ratio}
\end{figure}

We divided the 4$-$30~keV band into nine logarithmically spaced bins as listed in Table~\ref{tab:ebands}. In Fig.~\ref{fig:scox1} we show the radial surface-brightness profiles of {\sco} in E1$-$9 energy intervals. As expected, the shape and normalization of the {\art} {\spsf} at energies below 12~keV do not significantly change up to large off-axis angles. Above 12~keV, the {\spsf} normalization starts to decrease without a strong distortion of the shape of its energy dependence. To demonstrate the relative change of {\spsf} shape in different energy bands, we constructed the ratio of E2$-$E9 radial profiles to {\spsf} profile in 4.0$-$5.0~keV (E1) band (Fig.~\ref{fig:scox1:ratio}). 

\begin{figure}
\centerline{\includegraphics[width=\columnwidth]{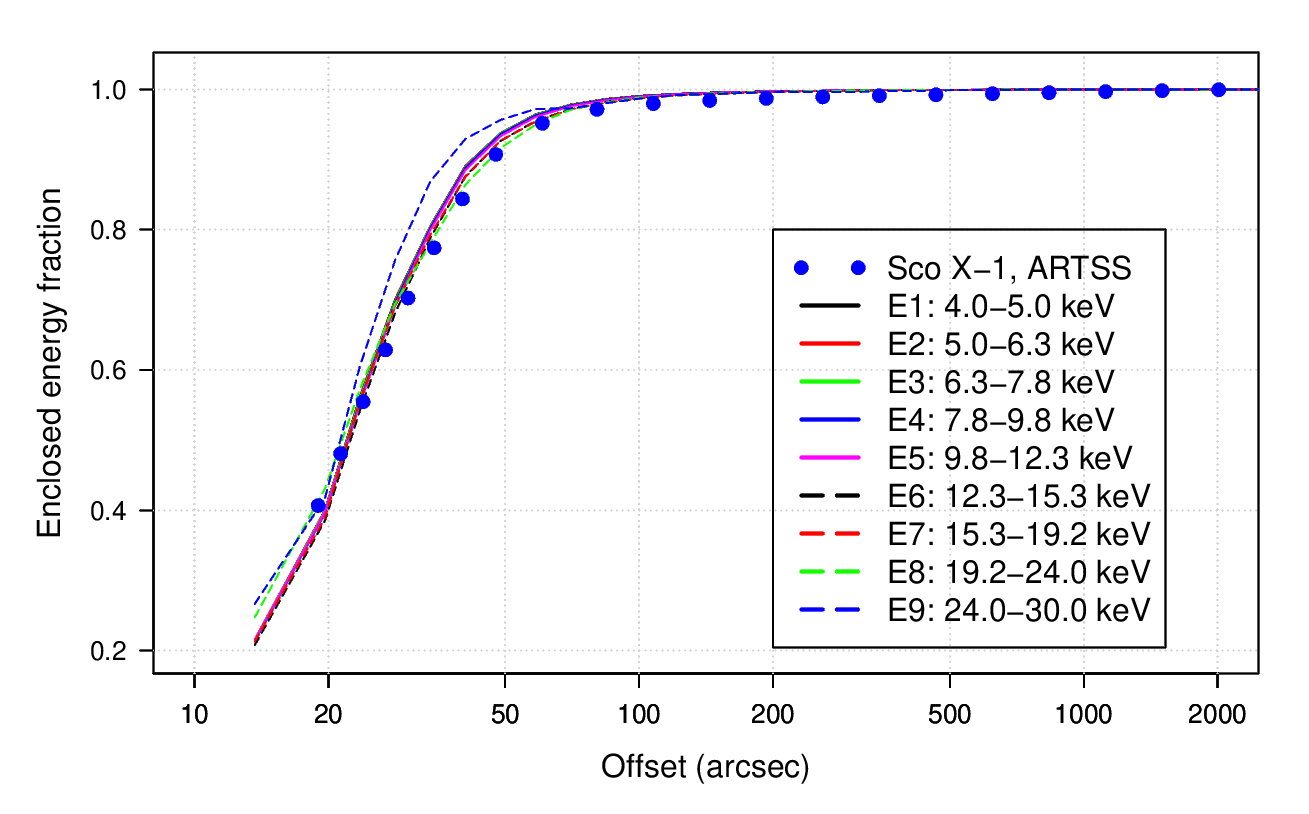}}
\centerline{\includegraphics[width=\columnwidth]{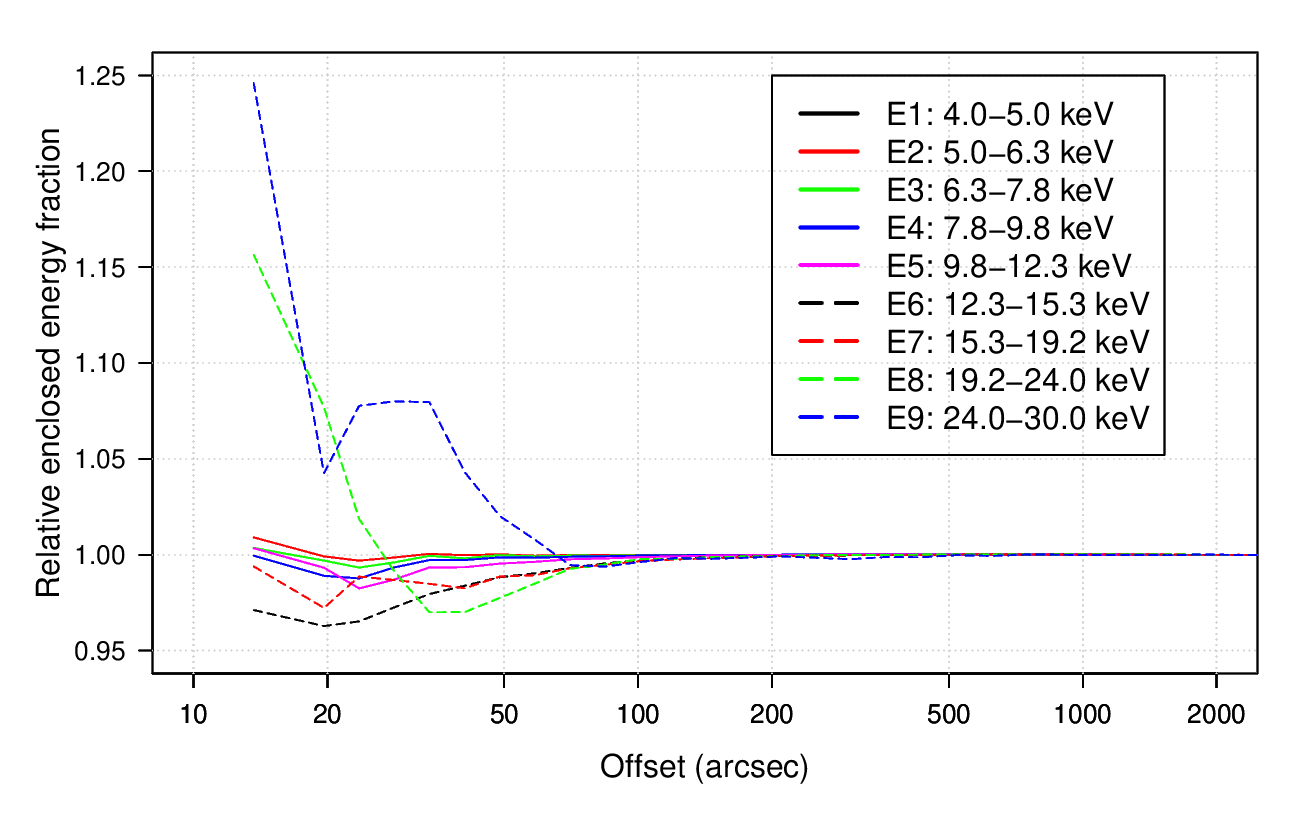}}
\caption{\textit{Upper panel}: Enclosed energy fraction (EEF) of the \art\ {\spsf} as a function of energy, based on {\sco} data. For comparison with Fig.~\ref{fig:eef_spfs} we added 4$-$12~keV band shown as a blue points ({\sco}, ARTSS). \textit{Bottom panel}: The same EEFs of E2$-$9 bands, shown as a ratio to 4.0$-$5.0~keV band (E1).  Note that the EEFs are similar within $\sim5\%$ of each other except for the highest energy values which differ by as much as $20\%$ at small off-axis angles.}
\label{fig:eef_spfs_energy}
\end{figure}

To illustrate how {\spsf} shape depends on energy, we constructed EEFs in the different energy bands E1$-$9, as shown in Fig.~\ref{fig:eef_spfs_energy} (upper panel). To compare with Fig.~\ref{fig:eef_spfs}, we also include the EEF in 4$-$12~keV band, also based on {\sco} data. Surprisingly, the different EEFs are almost superimposed on each other, except the E8 and E9 band (19.2$-$30~keV). A more detailed comparison of the relative variations of the EEFs is shown in the lower panel of Fig.~\ref{fig:eef_spfs_energy}, which reveals significant ${>}10\%$ deviation at lower off-axes angles ${<}20''$ for the E8$-$9 bands.  Since EEF, as a relative metric, is only sensitive to {\spsf} shape distortions, we conclude that the {\art} {\spsf} does not significantly change up to the end of reflectivity limit at 30~keV. Based on this, we conclude that the \art\ mirror PSF wings are dominated by purely geometric effects such as stray light caused by single-reflected photons. If the PSF wings were dominated by large-angle scatterings due to microroughness of the mirror surface, we would expect a strong energy dependence of the encircled energy fraction \citep[contrast with e.g.\ the situation with the Chandra PSF wings,][]{2004SPIE.5165..411G}, not observed at better than $\sim 5\%$ level at $E<20\,$keV.

\begin{figure}
\centerline{\includegraphics[width=\columnwidth]{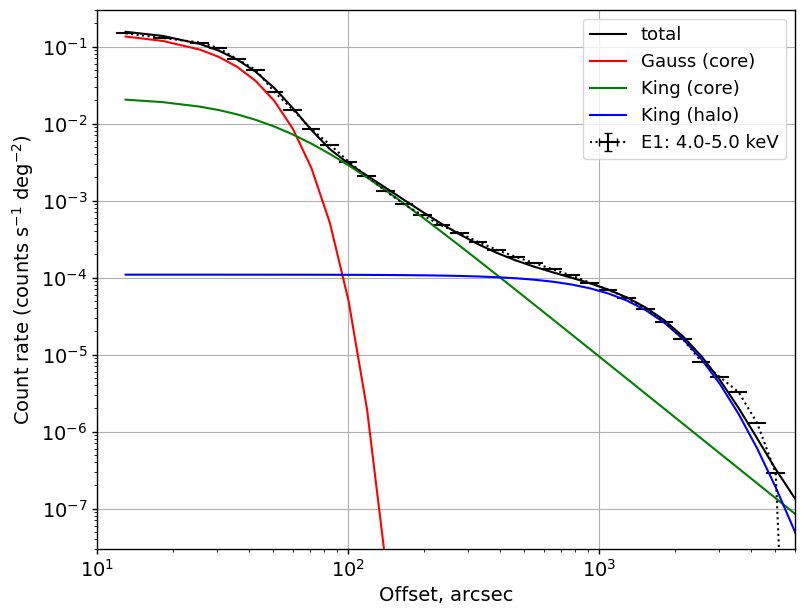}}
\caption{Best-fitting approximation of the {\art} {\spsf} in E1 (4$-$5~keV) energy band (Table~\ref{tab:ebands}) with three components, based on observations of {\sco} in survey mode.}
\label{fig:scox1:bestfit}
\end{figure}

We applied a parametrization of the {\spsf} needed for calibration purposes and many analytical studies. We found that the inflight {\spsf} is not described by two King functions (Sect.~\ref{sec:msfc}) due to the effective convolution with the large flight detector pixels which results in broadening of the central part of the {\spsf}. Instead, we describe the shape of the {\spsf} with a three-component model. The core is modeled by a Gaussian and a King function. The Gaussian,  $G(x)=N_{\rm G}\times Exp(-x^2/2\sigma_{\rm G}^2)$, where $x$ is the off-axis angle in arc-second, is fitted allowing the Gaussian width, $\sigma_{\rm G}$, and normalization, $N_{\rm G}$, free to vary. The core King function is approximated with the  normalization, $N_{\rm core}$, extent, $\sigma_{\rm core}$, and slope, $\gamma_{\rm core}\geq1$, free to vary in the fit. The halo King function is fitted with the normalization, $N_{\rm halo}$, and extent, $\sigma_{\rm halo}$, free to vary and the slope, $\gamma_{\rm halo}=5$, fixed. Note that in contrast to fitting {\art} PSF function measured on the ground (Sect.~\ref{sec:msfc}), all three components of {\spsf} now have independent normalizations. In Fig.~\ref{fig:scox1:bestfit} we show an example of the best-fitting approximation for the E1 energy band. The resulting parameters of the modeling for all energy ranges are listed in Table~\ref{tab:ebands}.

\begin{figure}
\centerline{\includegraphics[width=\columnwidth]{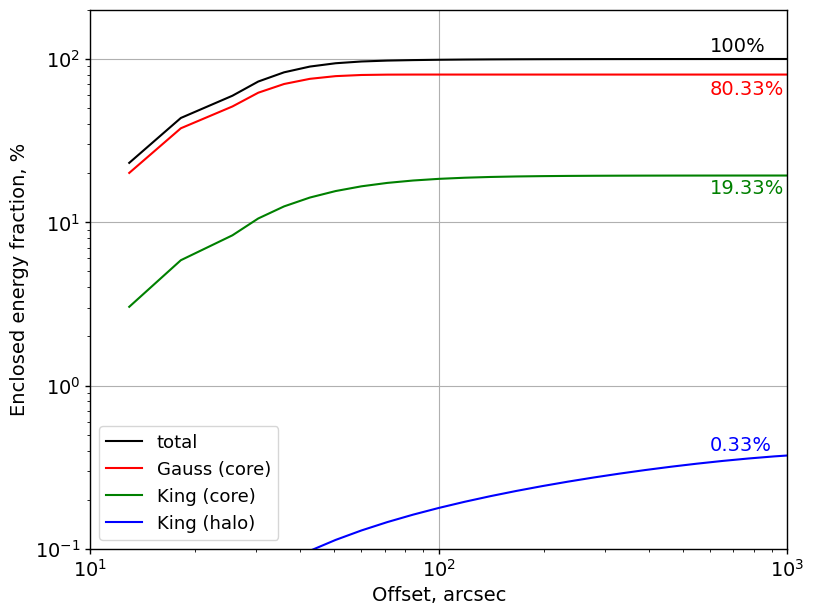}}
\caption{Enclosed energy fraction of different components of the {\art} {\spsf} in E1 (4$-$5~keV) energy band (Fig.~\ref{fig:scox1:bestfit}).}
\label{fig:scox1:eef}
\end{figure}

To estimate the relative contribution of different {\spsf} components, we show the EEF in Fig.~\ref{fig:scox1:eef}. The dominant fraction (${\sim}80\%$) is occupied by the Gaussian, and the rest is a sum of the two King components. The contribution of the widest King is less than 1\%.

\begin{sidewaystable}
\caption{Energy intervals used for the calibration of {\art} {\spsf} energy dependence and corresponding best-fitting parameters of approximation (Sect.~\ref{sec:discussion:energy}).}\label{tab:ebands}
\begin{tabular*}{\textheight}{@{\extracolsep\fill}ccccccccccccccccc}
\toprule%
& Range & $N_{\rm G}$ & $\sigma_{\rm G}$  & $N_{\rm core}$ & $\sigma_{\rm core}$ & $\gamma_{\rm core}$ & $N_{\rm halo}$ & $\sigma_{\rm halo}$  \\
& (keV) & (cts s$^{-1}$ deg$^{-2}$) & ($''$)  & (cts s$^{-1}$ deg$^{-2}$) & ($''$) &  & (cts s$^{-1}$ deg$^{-2}$) & ($''$)  \\
\midrule
E1 & 4.0$-$5.0&      $(1.54\pm0.02)* 10^{-1}$ &24.97 $\pm$ 0.09  &   $(2.2\pm0.2)* 10^{-2}$ &   52 $\pm$ 2  &   1.31 $\pm$ 0.01  &   $(1.09\pm0.01)* 10^{-4}$ &  $(3.14\pm0.01)* 10^{3}$  \\
E2& 5.0$-$6.3 &      $(1.99\pm0.01)* 10^{-1}$ &24.97 $\pm$ 0.08  &   $(2.2\pm0.1)* 10^{-2}$ &   64 $\pm$ 2  &   1.44 $\pm$ 0.01  &   $(1.59\pm0.01)* 10^{-4}$ &  $(2.92\pm0.01)* 10^{3}$  \\
E3& 6.3$-$7.8 &      $(2.05\pm0.01)* 10^{-1}$ &25.25 $\pm$ 0.08  &   $(1.5\pm0.1)* 10^{-2}$ &   85 $\pm$ 3  &   1.64 $\pm$ 0.02  &   $(1.88\pm0.01)* 10^{-4}$ &  $(2.63\pm0.01)* 10^{3}$  \\
E4& 7.8$-$9.8 &      $(1.63\pm0.01)* 10^{-1}$ &25.28 $\pm$ 0.09  &   $(1.2\pm0.1)* 10^{-2}$ &   95 $\pm$ 4  &   1.80 $\pm$ 0.04  &   $(1.68\pm0.01)* 10^{-4}$ &  $(2.27\pm0.01)* 10^{3}$ \\
E5& 9.8$-$12.3 &     $(1.01\pm0.01)* 10^{-1}$ &24.82 $\pm$ 0.14  &   $(8.3\pm0.1)* 10^{-3}$ &   86 $\pm$ 5  &   1.63 $\pm$ 0.04  &   $(1.03\pm0.01)* 10^{-4}$ &  $(2.12\pm0.01)* 10^{3}$ \\
E6& 12.3$-$15.3 &    $(4.82\pm0.03)* 10^{-2}$ &24.82 $\pm$ 0.35  &   $(1.1\pm0.3)* 10^{-2}$ &   57 $\pm$ 10  &   1.50 $\pm$ 0.07  &   $(8.65\pm0.03)* 10^{-5}$ &  $(1.81\pm0.02)* 10^{3}$  \\
E7& 15.3$-$19.2 &    $(2.44\pm0.02)* 10^{-2}$ &24.60 $\pm$ 0.79  &   $(3.8\pm2.2)* 10^{-3}$ &   71 $\pm$ 27 &   1.65 $\pm$ 0.30  &   $(5.16\pm0.05)* 10^{-5}$ &  $(1.62\pm0.04)* 10^{3}$  \\
E8& 19.2$-$24.0 &    $(3.09\pm0.02)* 10^{-3}$ &26.70 $\pm$ 6.02  &   $(8.0\pm3.9)* 10^{-3}$ &   32 $\pm$ 11 &   1.5 $\pm$ 0.3  &   $(1.59\pm0.05)* 10^{-5}$ &  $(1.68\pm0.20)* 10^{3}$ \\
E9& 24.0$-$30.0 &    $(3.17\pm0.01)* 10^{-3}$ &20.85 $\pm$ 3.05  &   $(6.6\pm0.2)* 10^{-5}$ &   80 $\pm$ 200  &   1.00   &   $(4.5\pm0.1)* 10^{-6}$ &  $(1.68\pm1.60)* 10^{3}$ \\
\botrule
\end{tabular*}
\end{sidewaystable}

\section{Summary}
\label{sec:summary}

The effective use of the {\art} telescope with FOV of $36'$ (in diameter) is achieved in scan or all-sky survey modes, which allow to cover large sky areas with unprecedented sensitivity. The knowledge of the telescope's X-ray mirror response plays a critical role in detecting weak point sources and studying extended objects with low surface brightness. However, the calibration of the PSF with real detector and at large offset angles is not always possible in the lab. In this work we calibrate the shape of the {\art} PSF at large angular distances using inflight observations of {\sco} and Crab Nebula.  

We demonstrated that the measurement of PSF in a combination with real detector pixel, is consistent with re-analysed ground calibration at MSFC. The shape of PSF, obtained with {\sco} observations can be used to model the wide wing of the {\art} PSF up to large angular distances. The estimated HPD of the {\art} PSF  in the scan and all-sky survey modes is about $48''$. The radial profile of the Crab Nebula in 4$-$12~keV band shows an extended structure at distances less than ${\sim}150''$, and at larger offsets, is  consistent with {\sco} data. Finally, using {\sco} data, we performed an analytic parametrization of the {\art} PSF as a function of energy in the nine intervals from 4 to 30 keV. 

The calibration of the PSF at large offsets helped to remove the contribution of bright point X-ray sources from the {\art} image of the Galactic Center, revealing the extended X-ray emission of the Nuclear Stellar Disk in 4$-$12~keV band. The shape of the {\art} PSF halo is also crucial for modeling the extended X-ray emission of the Coma cluster.

\bmhead{Acknowledgements}

The {\it Mikhail Pavlinsky} \art\ telescope is the hard X-ray instrument on board the \srg\ observatory, a flagship astrophysical project of the Russian Federal Space Program realized by the Russian Space Agency in the interests of the Russian Academy of Sciences. The \art\ team thanks the Russian Space Agency, Russian Academy of Sciences, and State Corporation Rosatom for the support of the \srg\ project and \art\ telescope. RK acknowledges support from the Russian Science Foundation (grant no. 24-22-00212).

\section*{Data Availability}
At the time of writing, the {\srg}/{\art} data and the corresponding data analysis software have a private status. We plan to provide public access to the {\art} scientific archive in the future.

\begin{appendices}
\end{appendices}


\bibliography{refs}

\end{document}